\documentclass{sig-alternate-05-2015}

\usepackage{xcolor}
\usepackage{listings}
\usepackage{subcaption}

\lstdefinestyle{base}{
  language=C,
  emptylines=1,
  breaklines=true,
  basicstyle=\ttfamily\scriptsize\color{black},
  moredelim=**[is][\color{red}]{@}{@},
  keywords={}
}

\newcommand{\fig}[1]{Figure~\ref{fig:#1}}

\newcommand{\sect}[1]{Section~\ref{sec:#1}}

\newcommand{\AX}{\emph{Access}}
\newcommand{\EX}{\emph{Execute}}

\begin{document}


\title{Decoupled Access-Execute on ARM big.LITTLE}
%
%
%
%
%
\numberofauthors{4} 
%
\author{
  \alignauthor
  Anton Weber \\
  \affaddr{Uppsala University}\\
  \email{anton.weber.0295\\@student.uu.se}
  \alignauthor
  Kim-Anh Tran \\
  \affaddr{Uppsala University}\\
  \email{kim-anh.tran\\@it.uu.se}
  \alignauthor
  Stefanos Kaxiras \\
  \affaddr{Uppsala University}\\
  \email{stefanos.kaxiras\\@it.uu.se}
  \and  
  \alignauthor
  Alexandra Jimborean\\
  \affaddr{Uppsala University}\\
  \email{alexandra.jimborean\\@it.uu.se}
}

\maketitle
\begin{abstract}
  Energy-efficiency plays a significant role given the battery lifetime
constraints in embedded systems and hand-held devices. In this work we
target the ARM big.LITTLE, a heterogeneous platform that is dominant
in the mobile and embedded market, which allows code to run
transparently on different microarchitectures with individual energy
and performance characteristics. 
It allows to use more energy efficient cores to conserve power during
simple tasks and idle times and switch over to faster, more power
hungry cores when performance is needed.

This proposal explores the power-savings and the performance gains
that can be achieved by utilizing the ARM big.LITTLE core in
combination with Decoupled Access-Execute (DAE). DAE is a compiler
technique that splits code regions into two distinct phases: a
memory-bound \AX{} phase and a compute-bound \EX{} phase. By
scheduling the memory-bound phase on the LITTLE core, and the
compute-bound phase on the big core, we conserve energy while caching
data from main memory and perform computations at maximum
performance. Our preliminary findings show that applying DAE on ARM
big.LITTLE has potential. By prefetching data in \AX{} we can achieve
an IPC improvement of up to 37\% in the \EX{} phase, and manage to
shift more than half of the program runtime to the LITTLE core. We
also provide insight into advantages and disadvantages of our
approach, present preliminary results and discuss potential solutions
to overcome locking overhead.
  
\end{abstract}


\pagebreak

\section{Introduction}
\label{sec:introduction}

Designed for embedded devices with strict size and energy constraints
\cite{sloss2004arm}, ARM was quickly adopted in modern mobile
hardware, such as smartphones and tablets, making it the de-facto
standard in these devices today. One of the main reasons for this is
the low-power RISC design, allowing for small, energy efficient chips
that at the same time are powerful enough to run modern mobile
operating systems.

While chip designs have been able to keep up with the rapid
development of the mobile segment in the past, ARM, just like its
competitors, is facing new challenges with the increasing demand on
performance and battery life in portable devices and the end of
Dennard scaling where it is no longer possible to lower transistor
size and keep the same power density. To address this issue, ARM
developed a new heterogeneous design named big.LITTLE that aims at
combining different CPU designs on the same system on chip (SoC).
The ARM big.LITTLE opens up
opportunities for the compiler to make scheduling decisions for phases
of the program with different performance characteristics. One such
compiler technique that can profit from the processor set up in the
ARM big.LITTLE is Decoupled Access-Execute
(DAE)\cite{jimborean2014fix}.

DAE has been developed as a way to improve performance and energy
efficiency on modern CPUs. It decouples code in coarse-grained
phases, namely memory-bound \AX{} phases  and
 compute-heavy  \EX{} phases. Using Dynamic
Voltage Frequency Scaling (DVFS) these phases can be executed at
different CPU frequencies and voltages. During the memory-bound phase
the CPU is stalled waiting for data to arrive from memory and can be
clocked down to conserve energy without performance loss. Once the
code reaches the \EX{} phase, the CPU can be scaled back to perform
the compute-heavy tasks at maximum performance. As a result,
processors can save energy by executing parts of a program at lower
voltage or frequency without slowing down the overall execution.

In this work we bring the two concepts together, not only to
demonstrate the advantages of Decoupled Access-Execute on ARM, but
also to adapt the previous ideas to the new architecture and take
advantage of the unique features that the heterogeneous design of ARM
big.LITTLE has to offer. In particular, we want to benefit from two
different core designs with individual performance and energy
characteristics being available to run decoupled execution. To this end,
our contributions are:

\begin{description}
\item[1.DAE on ARM big.LITTLE] We propose a new transformation pattern
  that shows how DAE can be applied to efficiently use the big and
  LITTLE cores of the ARM big.LITTLE architecture.
\item[2.Proof-of-Concept Implementation] We provide a proof-of-concept
  implementation of our transformation patterns for a selection of
  benchmarks. Our ideas lay the ground for the development of
  automatic compiler transformations for DAE on ARM big.LITTLE.
\end{description}

Our experiments show that \EX{} phases can significantly benefit from
prefetched data from \AX{} phases. By scheduling memory-bound phases
on \AX{} and compute-bound phases on \EX{}, we effectively reduce the
time spend on the big core (we observe IPC improvements of up to 37\%
for analyzed benchmarks), with more than half of the runtime being
spent on the LITTLE core. While being a proof-of-concept, our current
implementation still introduces a big synchronization overhead.
However, we present solutions to overcome it.


\section{Background}
\label{sec:background}

In the following, we will first give an overview on existing DAE
transformations. Afterwards, we will introduce details on the
processor in focus, the ARM big.LITTLE.

\subsection{Decoupled Access-Execute}
\label{sec:decoupled-access-execute}
While caches and hardware prefetchers have been introduced to decrease latency when accessing data from main memory, it still remains a common bottleneck in current computer architectures. With the processor waiting for data to arrive, there is not only a negative impact on the overall runtime of programs, but also on energy consumption.

One reason for this is that the processor runs at high(est) frequency while stalled in order to compute at full performance as soon as the data arrives. Lowering the frequency to reduce energy consumption during stalls would result in slower computations and further increase in program execution time. As an ideal approach, the memory-bound instructions that stall the CPU should be executed at low frequencies while the processor should run at maximum frequency for compute-bound parts of the program. 

Spiliopoulos et al.~\cite{spiliopoulos2011green} explored this possibility and created a framework that detects the memory- and compute-bound regions in the program and scales the CPU's frequency accordingly. Their work shows that this is indeed a viable approach but only if the regions are coarse enough. 

One of the reasons for this is that using current techniques, such as DVFS, switching the frequency or core voltage on the processor involves a significant transition overhead, preventing us from just switching between low and high frequencies as rapidly, as the ideal approach would require.

With Decoupled Access-Execute (DAE), Koukos et al.~\cite{koukos2013towards} proposed a solution to this problem by grouping memory- and compute-bound instructions in a program, creating larger code regions and thus reducing the number of frequency switches required. The result are two distinct phases, referred to as \AX{} (memory-bound) and \EX{} phase (compute-bound). 

Jimborean et al.~\cite{jimborean2014fix} created a set of compiler passes in LLVM allowing for these transformations of the program to be performed statically at compile-time. These passes create the \AX{} phase by duplicating the original code and stripping the copy of all instructions that are not memory reads or address computations. The original code becomes the \EX{} phase.

While the added \AX{} phase introduces overhead initially, the \EX{} phase will no longer be stalled by memory accesses, as the data will be available in the cache. In addition, the \AX{} phase allows for more memory level parallelism, potentially speeding up main memory accesses compared to the original code.

\begin{figure}[!htbp]
\centering
\includegraphics[width=0.3\textwidth]{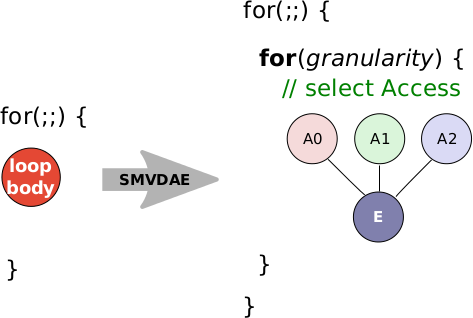}
\caption{Loop transformation with multiple \AX{} phases in SMVDAE.}
\label{fig:loop_chunking}
\end{figure}

Software Multiversioned Decoupled Access-Execute (SMVDAE)~\cite{koukos2016multiversioned} deals with the difficulties of applying DAE on general purpose applications. Here complex control flow and pointer aliasing make it hard to statically reason about the efficiency of the transformations. The SMVDAE compiler passes apply the transformations with a variety of parameters, thus generating multiple different access phase candidates, that allow to find the best performing options for the given code dynamically. With this approach, Koukos et al. were able to achieve on average over 20\% energy-delay-product (EDP) improvements across 14 different benchmarks.

To create coarse grained code regions, DAE targets hot loops within programs. These loops are split into smaller \textit{slices} that typically consist of several iterations of the original loop. \fig{loop_chunking} illustrates the SMVDAE transformation: for each slice of the original code, SMVDAE creates different \AX{} phase versions and one \EX{} phase. Using a dynamic version selector, the best performing \AX{} version is selected during runtime. As the amount of data accessed in each loop iteration depends on the task, the optimal size for the \textit{slices}, the \textit{granularity}, is benchmark- and cache dependent.

\subsection{ARM big.LITTLE}
Since its first prototype, the ARM1 from 1985, many iterations of ARM CPU cores have been developed. One key characteristic that distinguishes all these core designs from competitors like Intel and AMD is the RISC architecture. Instead of implementing a set of complex instructions in hardware, the RISC architecture shifts the complexity away from the chip and towards the software~\cite{sloss2004arm}. This makes RISC compiler tools more complex but also allow for much simpler hardware designs on the CPU. As a result, ARM has been able to create small, energy efficient chips that have seen great popularity in the embedded and mobile segment where size and battery life are major factors.

A recent development by ARM is the big.LITTLE architecture. First released in 2011, this heterogeneous design combines fast, powerful cores with slower, more energy efficient ones on the same SoC, allowing the devices to conserve power during small tasks and idle states but at the same time deliver high performance when needed. While the big and LITTLE processor cores are fully compatible from an instruction set architecture (ISA) level, they feature different microarchitectures. big cores are characterized by complex, out-of-order designs with many pipeline stages while the LITTLE cores are more simple, in-order processors. In comparison, the ARM Cortex-A15 (big) pipeline has 15 integer and 17-25 floating point pipeline stages, while the Cortex-A7 (LITTLE) only has 8 pipeline stages.

In modern ARM SoCs, CPU cores are grouped in \textit{clusters}. Each core has an individual L1 data and instruction cache but shares the L2 cache with other cores in the same cluster. CPU clusters and other components on the chip, such as GPU and peripherals, are connected through a shared bus. ARM provides reference designs for these interconnects, but manufacturers commonly use custom implementations in their products. The interconnect uses ARM's Advanced Microcontroller Bus Architecture (AMBA) and provides system-wide coherency through the AMBA AXI Coherency Extensions (ACE) and AMBA AXI Coherency Extensions Lite (ACE-Lite)~\cite{arm2013ambaace}. These protocols allow memory coherency across CPUs, one of the main prerequisites for big.LITTLE processing~\cite{arm2013biglittle}.

ARM big.LITTLE usually features two clusters: one for all big cores and one cluster containing all LITTLE cores. These designs allow three different techniques for runtime migrations of tasks: cluster switching, CPU migration and Global Task Scheduling (GTS)~\cite{arm2013biglittle, chungheterogeneous}.

Cluster switching was the first and most simple implementation. Only one of the two clusters is active at a time while the other is powered down.  If a switch is triggered, the other cluster is powered up, all tasks are migrated and the previously used cluster is deactivated until the next switch.
In CPU migration, each individual big core is paired with a LITTLE core. Each pair is visible as one virtual core and the system can transparently move a task between the two physical cores without affecting any of the other pairs.
GTS is the most flexible method, as all cores are visible to the system and can all be active at the same time.

Cluster switching and GTS also allow for asymmetric designs, where the number of big and LITTLE cores does not necessarily have to be equal.

\begin{figure}
\centering
\includegraphics[width=0.25\textwidth]{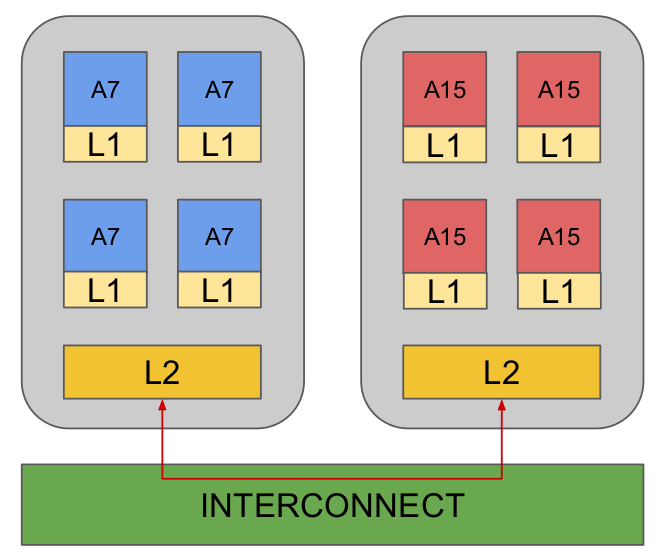}
\caption{Typical ARM big.LITTLE design}
\label{fig:big_little_overview}
\end{figure}


\section{Methodology}
\label{sec:methodology}

Our methodology applies DAE on ARM and takes advantage of the heterogeneous hardware features on the big.LITTLE architecture. With two  different types of processors available on the same system, energy savings can now be a result of lower core frequencies and running code on the simpler, more energy efficient microarchitecture of the LITTLE cores. A straight-forward way to benefit from this in decoupled execution, is to place our \AX{} and \EX{} phase onto the different cores.

Running the two phases on several CPUs also eliminates DVFS transition overhead, as the cores can constantly be kept at ideal frequencies throughout the entire execution. Since the cores are located on different clusters, we are no longer prefetching the data into a shared cache and are instead providing the \EX{} phase with prefetched data through coherence.

We have chosen to implement this approach for a selection of benchmarks.

\begin{figure}[!ht]
\centering
\includegraphics[width=0.25\textwidth]{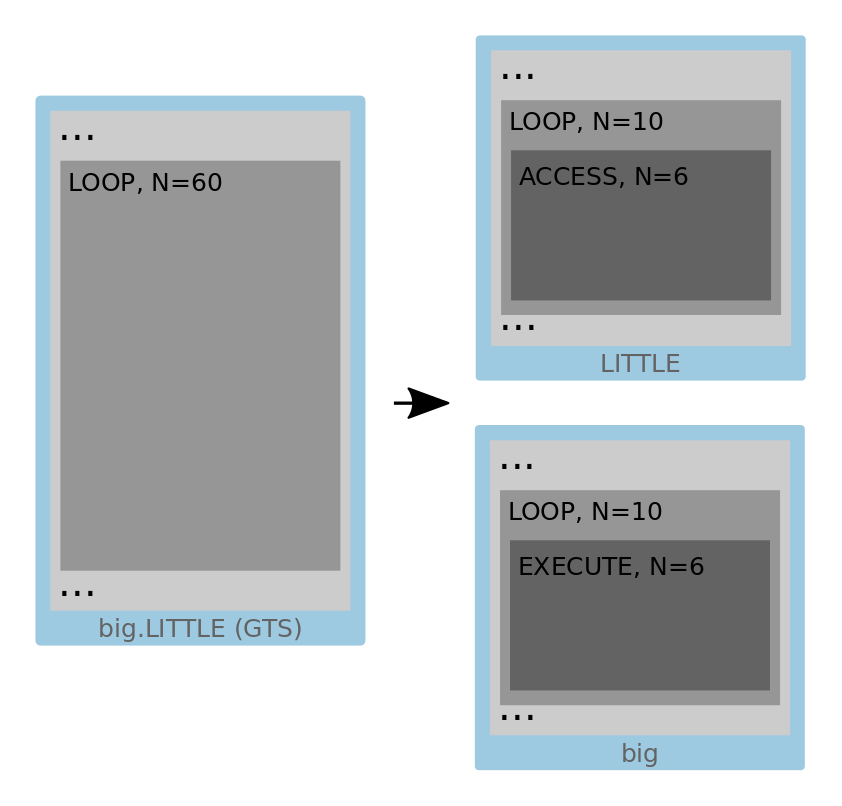}
\caption{Applying DAE techniques on big.LITTLE. The memory-bound \AX{}
  phase runs on the LITTLE core, while the compute-bound \EX{} phase
  runs on the big core.}
\label{fig:loop_chunking_mtdae}
\end{figure}

\subsection{Transformation pattern}
As our future goal is to have a set of compiler passes that can transform any input program, we design our new DAE transformations as a step by step process. The following sections describe this pattern and illustrate it on a sample program in pseudocode. The sample program in \fig{pattern-original} is used as a running example.

Although some of the steps have remained largely unchanged from current DAE compiler passes, we have chosen to implement this approach as prototypes in C code first. This allows us to adjust parameters and details in the implementation more rapidly and with more flexibility. 

\begin{figure}[!ht]
\centering
\begin{lstlisting}[style=base]
void do_work() {    //Main thread
  for(i=0;i<N;i++){
    c[i] = a[i+1]+b[i+2]
  }
}
\end{lstlisting}
\caption{Our running example: the original program. This example is kept small for the sake of simplicity.}
\label{fig:pattern-original}
\end{figure}

\subsubsection*{Spawning threads for Access and Execute phase}
Similar to the current DAE compiler passes, we are targeting hot loops. The loop is duplicated, but instead of executing the \AX{} and \EX{} phases within the same thread, we move them to individual cores (see \fig{loop_chunking_mtdae}). This is done by creating two threads: one for the \AX{}- and one for the \EX{} phase. 

The threads are created using the Linux POSIX thread interfaces (Pthreads). Configuring the attributes to the Pthread calls enables us to manually define CPU affinity and spawn the \EX{} phase on a big core and the \AX{} phase on a LITTLE core. This allows us to benefit from the flexibility of Global Task Scheduling and as our phases are meant to remain on the same core for the entire execution, we can avoid task migration and any of the associated overhead entirely. Spawning two additional threads instead of reusing the main thread for one of the phases allows us to run the phases on the two different cores without affecting the CPU affinity of the remaining parts of the program running on the main thread.

For the first transformation step this effectively means copying the loop twice and placing it into an empty function each. These two new functions will become our \AX{} and \EX{} phase. The original loop is replaced with the calls required to spawn two threads, one for each of the new functions, and join them when they finish. \fig{pattern-threading} shows the output of applying this transformation to the example program. The relevant changes are highlighted in red.

\begin{figure}[!ht]
\centering
\begin{lstlisting}[style=base]
void do_work() {    //Main thread
@  spawn(access_thread, access)@
@  spawn(execute_thread, execute)@
@  join(access_thread)@
@  join(execute_thread)@
}

@void access() {@    //Access thread
  for(i=0;i<N;i++){
    c[i] = a[i+1]+b[i+2]
  }
@}@

@void execute() {@    //Execute thread
  for(i=0;i<N;i++){
    c[i] = a[i+1]+b[i+2]
  }
@}@
\end{lstlisting}
\caption{We extract the original loop into a function (\EX{} phase)
  and duplicate the function to become the \AX{} phase. These two
  versions are executed as individual threads: one running on the LITTLE core,
  one on the big core.}
\label{fig:pattern-threading}
\end{figure}

\subsubsection*{Generating Access and Execute Phases}
Similar to previous DAE approaches, our \AX{} phases include control flow, loads and memory address calculations. Once all irrelevant instructions have been removed from the \AX{} phase, we can potentially optimize the resulting code during the compilation step, such as removing dead code or control flow that is no longer needed as a result of our changes.

The \EX{} phase remains unchanged after loop chunking (generation of slices) as we can issue the same data requests to the interconnect as the original code. All accesses to data that has already been prefetched by the \AX{} phase automatically benefit from it being available closer in the memory hierarchy, as the request will be serviced by the other cluster transparently.

Both phases contain address calculations. This means that we are now calculating addresses twice, introducing additional instructions compared to the original program. As in previous DAE implementations, we aim at compensating for this overhead through the positive side-effects of decoupled execution.

\fig{pattern-chunking} shows the generation of \AX{} and \EX{}: first we chunk the loop (i.e. creating an inner and outer loop), then we remove unnecessary instructions for \AX{} and replace loads with prefetch instructions.

\begin{figure}[!ht]
\centering
\begin{lstlisting}[style=base]
void do_work() {    //Main thread
  spawn(access_thread, access)
  spawn(execute_thread, execute)
  join(access_thread)
  join(execute_thread)
}

void access() {    //Access thread
  //Outer loop
  @offset=0@
  @for(j=0;j<(N/granularity);j++){@
    //Inner loop
    @for(k=0;k<granularity;k++){@
      @i=offset+k@
      @prefetch(a[i+1])@
      @prefetch(b[i+2])@
    }
    @offset+=granularity@
  }
}

void execute() {    //Execute thread
  //Outer loop
  @offset=0@
  @for(j=0;j<(N/granularity);j++){@
    //Inner loop
    @for(k=0;k<granularity;k++){@
      @i=offset+k@
      c[i]=a[i+1]+b[i+2]
    }
    @offset+=granularity@
  }
}
\end{lstlisting}
\caption{\AX{} Phase Generation: Chunking the loop and removing instructions from the \AX{} phase.}
\label{fig:pattern-chunking}
\end{figure}

\subsubsection*{Synchronization}
\begin{figure}
\includegraphics[width=0.5\textwidth]{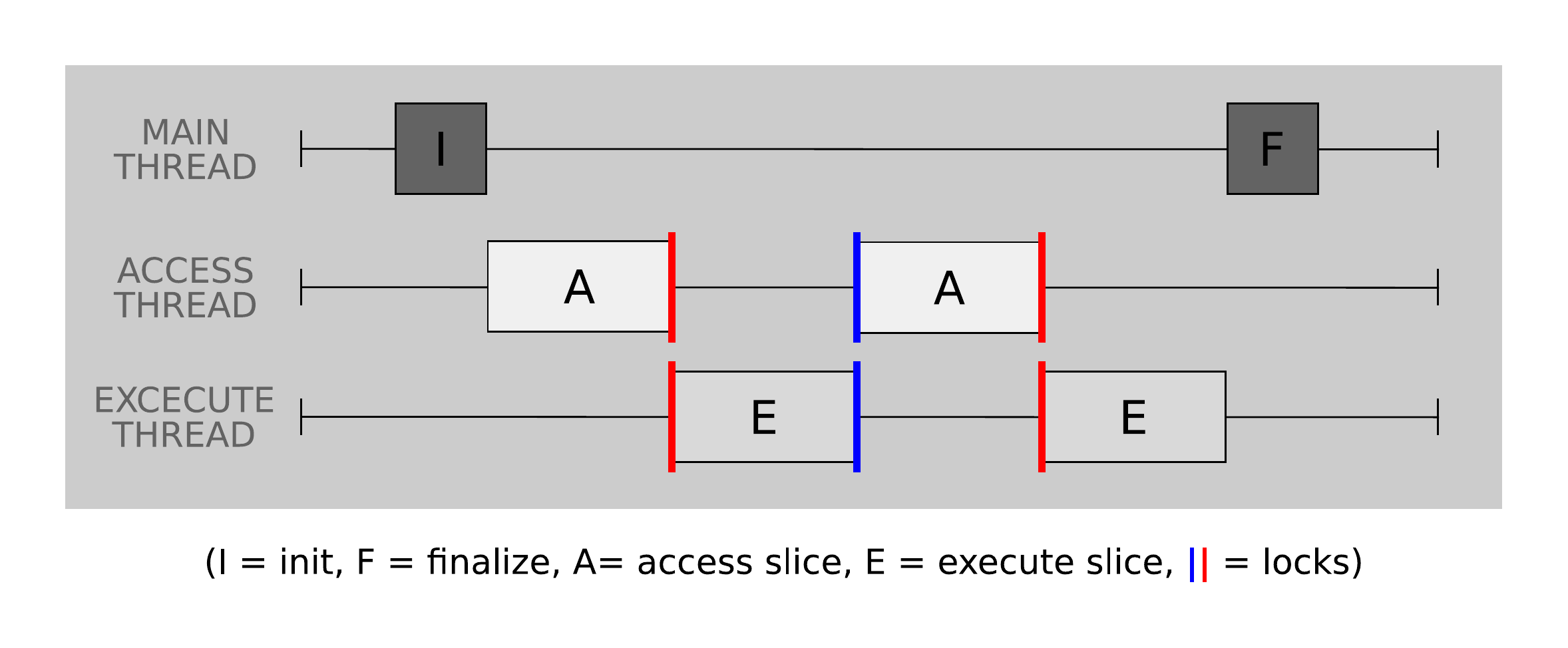}
\caption{Synchronization between individual \AX{} and \EX{} phases.}
\label{fig:mtdae-overview}
\end{figure}

As we are parallelizing decoupled execution, synchronization is required to enforce two rules: First, the \EX{} phase must not start computing before prefetching has finished, as it can only benefit from decoupled execution when the data is available in the cache as it is requested. And second, the \AX{} phase can not start the next slice before the \EX{} phase has completed the current one. As cache space is limited, prefetching the next set of data will potentially evict previously prefetched cache lines. As seen in \fig{mtdae-overview}, running \AX{} and \EX{} phases in turn on the two cores can be achieved using a combination of two locks. 

\fig{pattern-synchronization} shows how we can implement such a locking scheme in this transformation step. Each phase waits on one of these locks before starting the next chunk. 

\subsubsection*{Access lock (blue):} The first lock is used to start the \AX{} phase and unlocked at the end of each slice in the \EX{} phase. It is initially unlocked, so that the \AX{} phase can start as soon as the thread spawns. 

\subsubsection*{Execute lock (red):} The second lock signals the \EX{} phase to continue once the \AX{} phase finished the current slice. The \EX{} phase follows the same pattern with the difference that its first lock is initialized as locked, making sure that the phase does not start before the first \AX{} slice has been completed.

\begin{figure}
\centering
\begin{lstlisting}[style=base]
void do_work() {    //Main thread
@  init(access_lock)@
@  init(execute_lock)@
@  unlock(access_lock)@
@  lock(execute_lock)@
  spawn(access_thread, access)
  spawn(execute_thread, execute)
  join(access_thread)
  join(execute_thread)
}

void access() {    //Access thread
  //Outer loop
  offset=0
  for(j=0;j<(N/granularity);j++){
    //Inner loop
    @lock(access_lock)@
    for(k=0;k<granularity;k++){
      i=offset+k
      prefetch(a[i+1])
      prefetch(b[i+2])
    }
    offset+=granularity
    @unlock(execute_lock)@
  }
}

void execute() {    //Execute thread
  //Outer loop
  offset=0
  for(j=0;j<(N/granularity);j++){
    //Inner loop    
    @lock(execute_lock)@
    for(k=0;k<granularity;k++){
      i=offset+k
      c[i]=a[i+1]+b[i+2]
    }
    offset+=granularity
    @unlock(access_lock)@
  }
}
\end{lstlisting}
\caption{Adding synchronization between \AX{} and \EX{} phase.}
\label{fig:pattern-synchronization}
\end{figure}

\subsection{Optimizations}

The pattern above describes a basic implementation of DAE on big.LITTLE. Further optimizations to this methodology can improve results significantly in some scenarios.

While only the first optimization has been used as part of this work, all of the methods below have been proven to benefit decoupled execution on big.LITTLE through initial testing.

\subsubsection*{Reducing thread overhead}
As we create a pair of threads every time a loop is executed, the overhead of setting up, spawning and joining the threads can degrade performance noticeably when the program executes the loop frequently. A solution to this is to keep the threads running over the course of the program and provide them with new data on every loop execution. This thread-pool approach does not come without a downside as the threads need to be signaled when new data is available and the next loop should be executed. In many cases the benefits outweigh the added overhead and we decide experimentally whether to apply this optimization to the chosen benchmarks.

\subsubsection*{Overlap}
Running the \AX{} and \EX{} phases in parallel introduces a new way to reduce overall execution time by overlapping the two phases for each individual slice. While it is problematic to start the \EX{} phase too early, as the data has not yet arrived in the cache of the LITTLE cluster, it does not have to wait for the entire slice to be prefetched. 

In other words, to benefit from prefetched data in the \EX{} phase, the prefetch has to be completed by the time a particular data set is needed. Hence, any prefetches for data that is required at a later stage can still be pending when the \EX{} phase starts as long as they complete by the time the data is requested.

\begin{figure}[!h]
\includegraphics[width=0.5\textwidth]{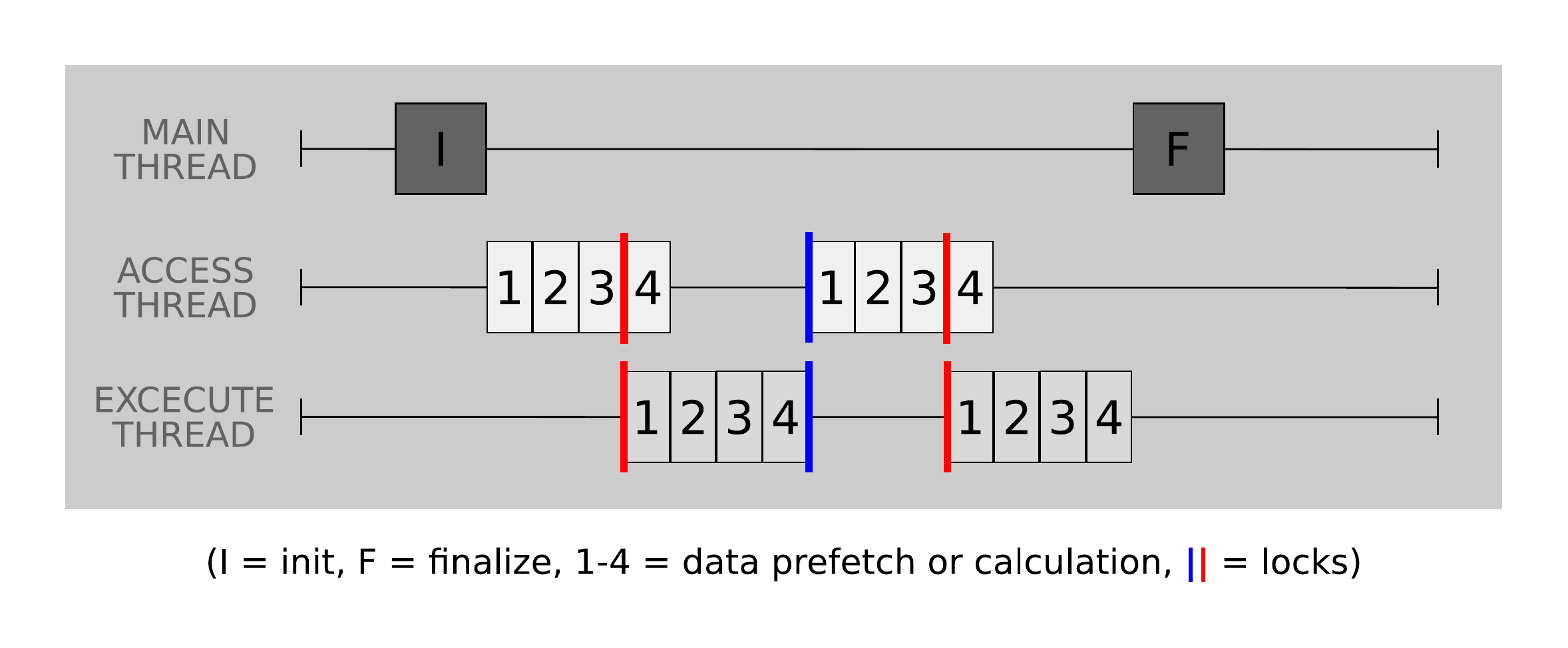}
\caption{Overlapping \AX{} and \EX{} phases.}
\label{fig:mtdae-overlap}
\end{figure}

\fig{mtdae-overlap} illustrates one possible scenario where the \EX{} phase is started before the last data set has been prefetched. As the data set is only required towards the end of the current \EX{} slice, the \AX{} phase can overlap with the early stages of the execute thread while \AX{} prefetches the last set of data in parallel.

Achieving the correct timing for this can be difficult, as the two phases are executed at different speeds and the \EX{} phase processes the data at a different rate than it is prefetched in the \AX{} phase.

In fact, as we  employ the Preload Data (PLD) instruction~\cite{arm2012architecture} to prefetch data in the \AX{} phase, we already deal with overlapping in the current implementation to some degree. The PLD prefetch hint is non-blocking, meaning that it will not wait for the data to actually arrive in the cache before completing. Hence, finishing the \AX{} phase only guarantees that we have issued all hints while any number of them might not have prefetched the data into the cache yet. We are currently extending this work to monitor the cache behavior (misses/hits for each big and LITTLE core) and origin of fetched data (local cache, cache of the other cores, memory).

\subsubsection{Timing-based implementation}
\label{sec:timing-based-implementation}
As we are expecting the synchronization overhead to be our main problem with the methodology described above, reducing or removing it entirely would greatly improve how the implementation performs.

While we generally need synchronization when dealing with multiple threads that work together, we have the advantage that we do not affect the correctness of the program by starting the \EX{} phase early or late. Incorrect timing merely affects the performance of the \EX{} phase, as the operations within the \AX{} phase are limited to address calculations an prefetches - i.e. operations that have no side-effects. As a matter of fact, we already have a loose synchronization model as a result of the non-blocking prefetches described in the section above.

\begin{figure}[!ht]
\includegraphics[width=0.5\textwidth]{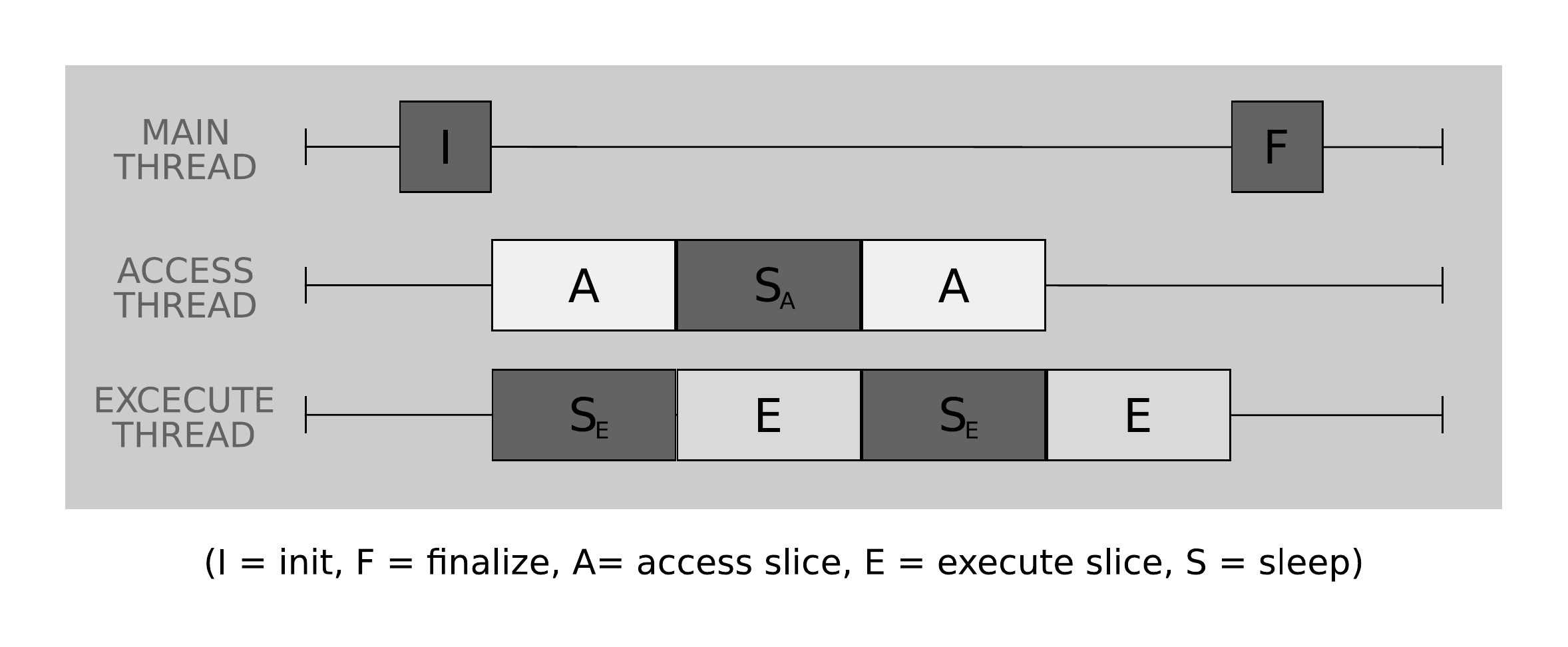}
\caption{Timing-based DAE: instead of locking, approximate the sleep
  times of \AX{} and \EX{}, in order to synchronize the phases.}
\label{fig:mtdae-lockfree}
\end{figure}

This allows us to go with a more speculative approach. Instead of locking, we can suspend the two threads and let the \AX{} and \EX{} phases wait for a specific amount of time before processing the next slice. The amount of wait time should correspond to the execution time of the other phase. Runtime measurements of the individual phases provide us with a starting point for these sleep times, while the exact values that perform best can be found experimentally. As a result, we can approximate a similar timing between the two threads without any of the locking overhead (see \fig{mtdae-lockfree}).

The main problem with this alternative is that it is difficult to reliably approximate the timing of the two threads as our system does not guarantee real-time constraints. Other programs running on the same system and the operating system itself can interfere with the DAE execution and affect the runtime of individual slices. Suspending the threads can be inaccurate, too, as using functions like \texttt{nanosleep} can be inaccurate and resume threads too late or return early.

The consequence in these cases is that the two threads can go out of sync, working on two different slices and negating all DAE benefits. The impact and likeliness of this to occur is dependent on many factors, including the number and size of the slices.

A trade-off would be a hybrid solution that only locks periodically to synchronize the two threads to the same slice. A group of slices would still be executed lock-free (i.e. with reduced overhead) and the synchronization point would prevent one of the phases from running too far ahead. In a more advanced implementation, these moments can also be used to adjust the sleep times of the individual threads dynamically to adapt to the current execution behavior caused by system load and other external factors.

\begin{figure}[!ht]
\centering
\begin{lstlisting}[style=base]
void execute() {    //Execute thread
  //Outer loop
  offset=0
  for(j=0;j<(N/granularity);j++){
    //Inner loop    
    @nanosleep(S)@
    for(k=0;k<granularity;k++){
      i=offset+k
      c[i]=a[i+1]+b[i+2]
    }
    offset+=granularity
  }
}
\end{lstlisting}
\caption{Replacing both locks with a single sleep call in each phase.}
\label{fig:pattern-timing}
\end{figure}


\section{Evaluation}
\label{sec:evaluation}

In this section we first describe the ARM big.LITTLE that we use in
our evaluation. We further describe our measurement techniques and our
evaluation criteria. Afterwards we present and discuss the
experimental results obtained on the selected benchmarks.

\subsection{Experimental Setup}
\subsubsection*{Test system}
We evaluate the benchmarks on an ODROID-XU4 single-board computer, running the Samsung Exynos 5422 SoC. This ARMv7-A chip features one big and one LITTLE cluster. Each cluster shares the L2 cache while each individual core has a private L1 data and private L1 instruction cache available. The 2 GB of LPDDR3 main memory is specified with a bandwidth of 14.9 GB/s. Table \ref{table:exynos-specs} contains more details about the processors on this chip.

\begin{table}[!ht]
\centering
\begin{tabular}{ l | c | c }
   & big cluster & LITTLE cluster \\
  \hline	
  Number of cores & 4 & 4 \\
  Core type & Cortex-A15 & Cortex-A7 \\
  $f_{max}$ & 2 GHz & 1.4 GHz \\
  $f_{min}$ & 200 MHz & 200MHz \\
  L2 cache & 2MB & 512 kB \\
  L1d cache & 32kB &32 kB \\

\end{tabular}
\caption{Exynos 5422 specifications.}
\label{table:exynos-specs}
\end{table}

The operating system is a 64 bit Ubuntu Linux with a kernel maintained by ODROID that contains all relevant drivers to run the scheduler in GTS mode. To reduce scheduler interference, two processor cores are removed from the scheduler queues using the \textit{isolcpus} kernel option. The benchmarks are cross-compiled on an Intel-based x86 machine using LLVM 3.8~\cite{llvm}.

\subsubsection*{Measurement technique}
Measurements through the Linux \textit{perf\_events} interface, while convenient to set up, have shown to produce too much overhead - especially with frequent, fine-grained measurements. 
Instead, we directly access the performance statistics available as part of the on-chip Performance Monitor Units (PMUs). In addition to a dedicated cycle counter, there is a number of configurable counters available (4 per A7 core and 6 on the Cortex-A15~\cite{arm2013a7, arm2013a15}). One of these event counters on each processor is set to capture the number of retired instructions~\cite{arm2012architecture}. The cycle counters are configured to increment once every 64 clock cycles to avoid overflows in the 32 bit registers when measuring around larger code regions. We enable PMU counters and allow user-space access through a custom kernel module. This enables us to read the relevant register values through inline assembly instructions from within our code.

\subsection{Evaluation criteria}
As the goal of this project is improve energy efficiency of programs without sacrificing performance, these two aspects will be our main criteria for evaluating the prototypes.

Similar to Software Multiversioned Decoupled Access Execute (SMVDAE)~\cite{koukos2016multiversioned}, we generate multiple versions with different granularities for each benchmark to find the optimal setting.

\subsubsection*{Performance:}
We measure the performance impact with the overall runtime of the benchmark. While previous DAE work has shown to speed up benchmark execution in some cases, we now expect synchronization and coherence overhead to affect our results.

\subsubsection*{Energy:} 
Without a sophisticated power model, speeding up the code run on the big core, together with the individual execution times for each phase, are our main indicators in terms of energy savings. Speeding up the \EX{} phase means that the big core will be active for less time and hence consume less power. On the other hand, these savings are only relevant if the overall overhead is small enough not to negate this effect.

\subsubsection*{Measurement considerations:} 
Changing the original benchmark into a threaded version introduces synchronization overhead. This is reflected in the runtime measurements, where faster execution does not only result from data being available faster, but also from lower synchronization overhead. This means that we need different measurements to evaluate how much we speed up the \EX{} phase purely by providing the data from our warmed up cache.

Currently, the most accurate way to measure this is to see how many CPU cycles the calculations in the execute phase need to finish. If the data is not available in cache, the instructions will take longer time to execute. On the other hand, if the data can be brought in through coherence, the core will spend less time stalled waiting for it to arrive and finish the instructions in fewer cycles.
The baseline for runtime comparisons is the unmodified version of each benchmark. While loop chunking alone can affect performance, the impact has shown to be negligible for the benchmarks we evaluate.

A common unit to visualize these results is instructions per cycle (IPC). For this we capture the cycle and instruction count in the \EX{} phase individually and calculate the IPC. The samples are taken around the inner loop of the \EX{} phase (see \fig{measurement-inner-loop}) to avoid capturing the execution of the other phase and locking overhead. These are measured separately. The IPC of the baseline is obtained by chunking the unmodified version of the benchmark, and by measuring the instuctions and cycles of the otherwise unchanged inner loop. Thus, the IPC numbers of both versions refer to the exact same region of code.

\begin{figure}[!ht]
\centering
\begin{lstlisting}[style=base]
void execute() {    //Execute thread
  ...
  //Outer loop
  for(j=0;j<(N/granularity);j++){
    lock(execute_lock)
    @start = read_counter()@
    //Inner loop 
    for(k=0;k<granularity;k++){
      ...
    }
    @end = read_counter()@
    ...
    unlock(access_lock)
  }
}
\end{lstlisting}
\caption{Measuring the effect of \AX{} on \EX{}: we take measurements
  around the inner loop, in order to avoid capturing overhead associated
  with locking or other phases.}
\label{fig:measurement-inner-loop}
\end{figure}

\subsubsection*{Benchmarks:}
For the evaluation of our approach we have modified two benchmarks from the SPEC 2006 suite~\cite{Henning06}, libquantum and LBM, and CIGAR~\cite{cigar}, a genetic algorithm search. All three benchmarks have been considered by Koukos et al.~\cite{koukos2013towards} in the initial task-based DAE framework and allow a comparison of DAE on big.LITTLE to previous results.

Each benchmark has individual characteristics and memory access patterns which have an impact on how the prototypes perform. While libquantum and CIGAR are considered memory-bound, LBM has been classified as intermediate~\cite{koukos2013towards}.

\subsection{Benchmark results}

\subsubsection*{LBM}

The loop we target in LBM performs a number of irregular memory accesses with limited control flow. The if-conditions only affect how calculations are performed in the \EX{} phase while the required data remains the same in all cases. This results in a simplified \AX{} phase without any control flow, as we can always prefetch the same values, independent of which path in the control flow graph the \EX{} phase takes. Most of the values that are accessed from memory are double-precision floating point numbers.

The irregular access pattern and memory-bound calculations are an ideal target for decoupled execution. And in fact, the benchmark results show that we are able to improve the IPC of the \EX{} phase by up to 31\% by prefetching the data on the LITTLE core.

The total benchmark runtime increases significantly as we lower the granularity. A smaller granularity increases the number of total slices and with that also the number of synchronizations performed between the two phases. This additional locking overhead results in a penalty to overall runtime. The loop is executed several times based on an input parameter for the benchmark, which multiplies this negative effect.

\begin{figure}[!ht]
\centering
\includegraphics[width=0.5\textwidth]{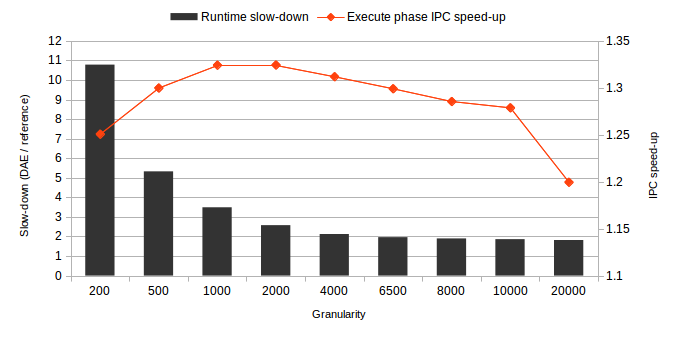}
\caption{\EX{} IPC and overall slow-down for LBM.}
\label{fig:lbm-results}
\end{figure}

\fig{lbm-results} relates these two findings, the IPC \textbf{speed-up} and the overall benchmark \textbf{slow-down}, to each other. While we observe the best runtime at large granularities, choosing a slightly higher overall slow-down results in much better \EX{} phase performance (i.e. we spend less time on the big core). This is discussed in more detail further below.

\subsubsection*{CIGAR}

In this case we apply DAE to a function with a higher degree of indirection in the memory accesses. The calculations themselves are relatively simple and compare two double values within a struct to determine a new maximum and swap two values within an integer array.

The results show that we can achieve slightly better IPC improvements compared to LBM with a peak speed-up of 37\%. The slow-down of overall execution time at lower granularities is still significant, yet not as extreme as in the previous benchmark. This can be explained by the smaller loop size (i.e. the same granularity results in less total slices) and the fact that the loop is only executed once as part of the benchmark.

The trend of the IPC graph indicated that we can potentially speed up the \EX{} phase even further by lowering the granularity. Yet, as the overhead at low granularities increases significantly, any improvement in IPC would be negated.

\begin{figure}[!ht]
\centering
\includegraphics[width=0.5\textwidth]{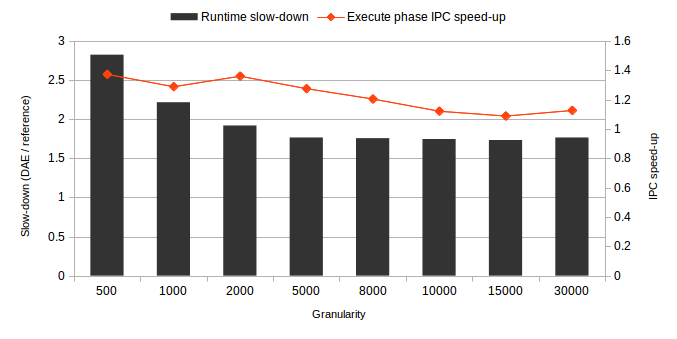}
\caption{\EX{} IPC and overall slow-down for CIGAR.}
\label{fig:cigar-results}
\end{figure}

\subsubsection*{libquantum}

While the other two benchmarks spawn threads every time they execute their loop, frequent calls to our target loop in libquantum make it a good candidate for the thread-pool optimization. As we in fact have been able to reduce the overhead noticeably in this case, we have taken all measurements with the thread-pool variant.

\begin{figure}[!ht]
\centering
\includegraphics[width=0.5\textwidth]{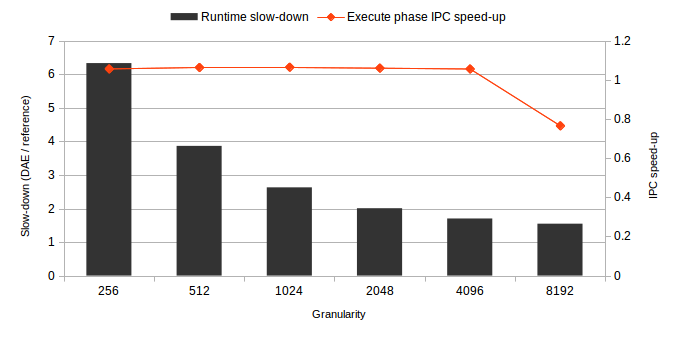}
\centering
\caption{\EX{} IPC and overall slow-down for libquantum.}
\label{fig:libq-results}
\end{figure}

The loop itself has a regular access pattern and performs a single bitwise XOR operation on a struct member on each iteration. Despite the memory-bound nature of this loop, we only observe a maximum \EX{} phase IPC speed-up of 6.7\% (see \fig{libq-results}). This is an interesting finding, as previous DAE evaluations of libquantum show improvements in energy and performance~\cite{jimborean2014fix, koukos2013towards, koukos2016multiversioned}. Further investigations are needed to determine whether this new behavior is caused by coherence side-effects or other new factors introduced by DAE execution on big.LITTLE or whether the shift to the new architecture of the Cortex-A15 and Cortex-A7 alone are responsible.

\subsection{Performance}

\begin{figure}[!ht]
\centering
\begin{subfigure}[b]{4cm}
\includegraphics[width=4cm]{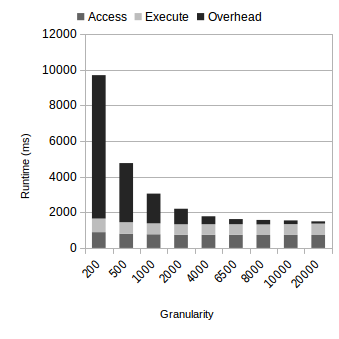}
\caption{LBM}
\label{Fig:breakdown-lbm}
\end{subfigure}
\begin{subfigure}[b]{4cm}
\centering
\includegraphics[width=4cm]{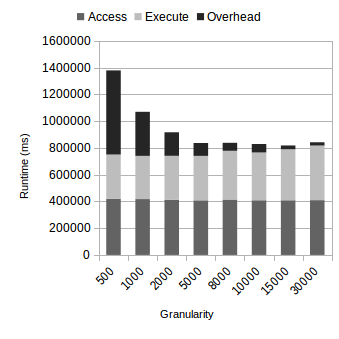}
\caption{CIGAR}
\label{Fig:breakdown-cigar}
\end{subfigure}
\hspace{0.5cm}
\begin{subfigure}[b]{4cm}
\centering
\includegraphics[width=4cm]{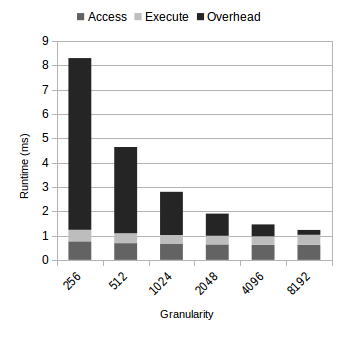}
\caption{Libquantum}
\label{Fig:breakdown-libq}
\end{subfigure}
\caption{Function runtime breakdown: runtime associated to \AX{}, \EX{} and synchronization.}\label{fig:breakdowns}
\end{figure}

Breaking down the time we spend inside the individual functions that contain the targeted loop, we can analyze which part (\AX{}, \EX{} or synchronization) is causing the increase in execution time. \fig{breakdowns} illustrates this for all three benchmarks. 

Here we can clearly see that, while the \EX{} phase gets faster, overhead is slowing down the overall execution as we reduce the granularity. This portion of the function time is dominated by the locking overhead between the two phases, as the time to initialize and join the threads is insignificant (\textless 1ms). As mentioned before, the locking overhead is proportional to the total number of slices, as each slice causes two locking operations. This makes large granularities perform significantly better. While the overhead of synchronizing \AX{} and \EX{} outweights benefits achieved from prefetching, the mechanism to synchronize can be exchanged by a lightweight mechanism in the future (such as the timing-based DAE described in \sect{timing-based-implementation}).

Compared to previous DAE implementations, our approach achieves less speed up for \EX{} phase. A major factor for this is the lack of a shared LLC on our system. Running the \AX{} phase on the A7 only brings in the data into the LITTLE cluster. As a result, touching prefetched data in the \EX{} phase no longer results in an instant cache hit. The cache miss is just serviced by the A7 cluster instead of by main memory. Effectively this means that we are not making data instantly available in the cache, but are merely reducing the cache miss latency on the A15.
The result is that we are now observing the full memory latency on the A7 and a reduced load time on the A15 instead of a much shorter combination of full memory latency in the \AX{} phase plus cache hit latency in the \EX{} phase.

\subsection{Energy savings}

The previous individual results show that we managed to speed up the \EX{} phase significantly for two out of the three benchmarks. While we were only able to do this at the cost of slowing down the overall execution time, it is important to consider that more than half of it is now spent on the LITTLE core as part of the \AX{} phase. During this time we not only run on a more energy-efficient microarchitecture but also at a lower base clock frequency with the option to lower it even further at relatively small performance penalties.

With excessive overhead potentially nullifying all energy savings that we achieve through speeding up the \EX{} phase, the current implementation of DAE for big.LITTLE requires us to find a balance between reducing the time we spend on the big core and keeping the overall runtime low. 
As we do not only observe notable IPC improvements for small granularities where locking overhead becomes problematic, but also for larger slices (e.g.~28\% IPC speed-up at 1.77x slow-down compared to 37\% peak improvement at 2.82x slow-down for CIGAR), the granularity effectively becomes main point of adjustment when we want to strike this balance.

The sub optimal performance that results from the lack of a shared LLC between the big and LITTLE core mentioned in the previous section also affects our energy consumption directly. Not being able to reduce the stalls in the \EX{} phase further means that we are spending more time on the power-hungry big core than an ideal DAE implementation would. On the other hand, running the protoype on a system with shared LLC, we expect to see significantly improved results in both performance and enery savings.


\section{Related Work}
\label{sec:related-work}

\subsection{Inter-core prefetching}
Kamruzzaman et al.~\cite{kamruzzaman2011inter} investigated how a multi-core system can be used to improve performance in single-threaded applications. Their \textit{inter-core prefetching} technique uses helper threads on different cores to prefetch data. The original compute thread is migrated between these cores at specific points in the execution to benefit from the data that has been brought into the caches. This technique performs transformations in software without the need of special hardware and, similar to the DAE approach, targets large loops within a program.

With memory-bound applications, they have been able to achieve up to 2.8x speed up and an average energy reduction between 11 and 26\%. It also shows the advantages of moving the prefetches to a different core in comparison to previous simultaneous multithreading (SMT) approaches. This prevents from competing for CPU resources with the main thread and avoids the negative impact on L1 cache behavior. On the other hand, it is also mentioned that this approach has downsides, such as problems with cache coherence when working on the same data in the main and helper threads.

While this concept has some similarities to the methodology we design for DAE on big.LITTLE, we avoid migrating tasks during execution and rely on accessing prefetched values through coherence. Additionally, we coordinate our \AX{} and \EX{} phases to work on the same chunk of data instead of prefetching ahead.

\subsection{big.LITTLE task migration}
\label{sec:big.LITTLE_task_migration}
big.LITTLE implementations currently choose between two distinct methods for task migration. The first method relies on CPU frequency frameworks, such as \textit{cpufreq}, and works with cluster switching and CPU migration. When a certain performance threshold is reached, tasks are migrated to another core or a cluster switch is triggered. This is comparable to traditional DVFS techniques with the difference that lower power is not represented by a change in CPU voltage or frequency but by a migration to a different core or cluster~\cite{arm2013biglittle}.

Global Task Scheduling relies on the OS scheduler to migrate tasks. In this model the scheduler is aware of the different characteristics of the cores in the system and creates and migrates tasks accordingly. For this, it tracks the performance requirements of threads and CPU load on the system. This data can then be used together with heuristics to decide the scheduling behavior~\cite{arm2013biglittle,chungheterogeneous}. As all cores are visible to the system and can be active at the same time, this method is regarded as the most flexible. Manufacturer white papers show that GTS improves benchmark performance by 20\% at similar power consumption compared to cluster switching on the same hardware~\cite{chungheterogeneous}.

\subsection{Scheduling on heterogeneous architectures}
Chen et al.~\cite{chen2009efficient} analyzed scheduling techniques on heterogeneous architectures. For this they created a model that bases scheduling decisions on matching the characteristics of the different hardware to the resource requirements of the tasks to be scheduled. In their work, they consider instruction-level parallelism (ILP), branch predictability, and data locality of a task and the hardware properties hardware issue width, branch predictor size and cache size. Using this method they reduce EDP by an average of 24.5\% and improve throughput and energy savings. While DAE does not consider the same workload characteristics, we are taking a related approach by matching the different phases to the appropriate core type within the heterogeneous big.LITTLE design.

Van Craeynest et al.~\cite{van2012scheduling} created a method to predict how the different cores in single-ISA heterogeneous architectures, including ARM big.LITTLE, perform for a given type of task and developed a dynamic scheduling mechanism based on their findings. Their work mentions that simple, in-order cores perform best on tasks with high ILP while complex out-of-order cores benefit from memory-level parallelism (MLP) or where ILP can be extracted dynamically. As a conclusion, choosing cores based on whether the execution is memory- or compute-intensive without taking those aspects into account can lead to sub optimal performance.

Decoupled execution focuses on the memory- or compute-bound properties to divide tasks into the two phases. While the work of Van Craeynest et al.~\cite{van2012scheduling} evaluates benchmarks as a whole and DAE is applied on a much finer scale within individual functions of the program, their insights are important to take into account when moving DAE to a heterogeneous architecture. When deciding which tasks to consider for the individual phases for example, taking the level of ILP and MLP into account becomes important as the phases can now be scheduled on different types of cores.


\section{Conclusions}
\label{sec:conclusion}
The results show that Decoupled Access-Execute on ARM big.LITTLE can indeed provide noticeable energy savings, but currently only at the expense of sacrificing performance due to the increased overhead of synchronization. Our prototypes successfully demonstrate decoupled execution on a heterogeneous architecture by running memory-bound sections on the energy-efficient LITTLE core and compute-bound parts of the program on the performance-focused big core. For two out of three benchmarks we were able to improve the \EX{} phase IPC significantly (up to 37\%), reducing the time they are executed at high frequencies and on performance-focused hardware.

As part of the development and evaluation process, we have identified the bottlenecks of the current implementation and suggest concrete optimization concepts for future iterations of this work. The locking overhead that has been introduced as a result of parallelizing the decoupled execution is the main reason for the overall benchmark slow-down. As it is proportional to the number of slices, choosing some granularities is no longer viable. Instead we are now facing a trade-off between IPC improvement and slow-down in the overall runtime with this approach. The proposed optimizations show the potential to reduce or remove this overhead entirely while still benefiting from decoupled execution.

While the synchronization overhead plays a big role, we are also limited by the fact that the CPU clusters in the Exynos 5422 do not share a LLC. For our implementation this has the far-reaching disadvantage that any prefetches performed in the A7 cluster are not directly available for the A15 cores and have to be brought in through coherence instead. This prevents DAE to perform at its full potential. While this problem is unavoidable on the current system, adding a shared LLC is a simple solution to it. In fact, newer interconnect designs used in recent homogeneous designs already support a shared L3 cache between clusters that is directly connected to the bus.



\bibliographystyle{abbrv}
\bibliography{refs}

\begin{thebibliography}{10}

\bibitem{llvm}
{The LLVM Compiler Infrastructure}, 2016.
\newblock [Online] accessed 2016-05-28. Available at \url{http://llvm.org/}.

\bibitem{arm2012architecture}
ARM.
\newblock {Architecture Reference Manual. ARMv7-A and ARMv7-R edition}.
\newblock 2012.

\bibitem{arm2013a7}
ARM.
\newblock {Cortex-A7 MPCore Processor Technical Reference Manual}.
\newblock 2013. Revision: r0p5.

\bibitem{arm2013a15}
ARM.
\newblock {Cortex-A15 MPCore Processor Technical Reference Manual}.
\newblock 2013. Revision: r4p0.

\bibitem{arm2013biglittle}
{ARM Limited}.
\newblock {big.LITTLE Technology: The Future of Mobile}.
\newblock Technical report, 2013.

\bibitem{chen2009efficient}
J.~Chen and L.~K. John.
\newblock Efficient program scheduling for heterogeneous multi-core processors.
\newblock In {\em Proceedings of the 46th Annual Design Automation Conference},
  pages 927--930. ACM, 2009.

\bibitem{chungheterogeneous}
H.~Chung, M.~Kang, and H.-D. Cho.
\newblock {Heterogeneous Multi-Processing Solution of Exynos 5 Octa with
  ARM{\textregistered} big.LITTLE{\texttrademark} Technology}.

\bibitem{Henning06}
J.~L. Henning.
\newblock {SPEC} {CPU2006} benchmark descriptions.
\newblock {\em {SIGARCH} Computer Architecture News}, 34(4):1--17, 2006.

\bibitem{jimborean2014fix}
A.~Jimborean, K.~Koukos, V.~Spiliopoulos, D.~Black-Schaffer, and S.~Kaxiras.
\newblock Fix the code. don't tweak the hardware: A new compiler approach to
  voltage-frequency scaling.
\newblock In {\em Proceedings of Annual IEEE/ACM International Symposium on
  Code Generation and Optimization}, page 262. ACM, 2014.

\bibitem{kamruzzaman2011inter}
M.~Kamruzzaman, S.~Swanson, and D.~M. Tullsen.
\newblock Inter-core prefetching for multicore processors using migrating
  helper threads.
\newblock In {\em ACM SIGARCH Computer Architecture News}, volume~39, pages
  393--404. ACM, 2011.

\bibitem{koukos2013towards}
K.~Koukos, D.~Black-Schaffer, V.~Spiliopoulos, and S.~Kaxiras.
\newblock Towards more efficient execution: A decoupled access-execute
  approach.
\newblock In {\em Proceedings of the 27th international ACM conference on
  International conference on supercomputing}, pages 253--262. ACM, 2013.

\bibitem{koukos2016multiversioned}
K.~Koukos, P.~Ekemark, G.~Zacharopoulos, V.~Spiliopoulos, S.~Kaxiras, and
  A.~Jimborean.
\newblock Multiversioned decoupled access-execute: the key to energy-efficient
  compilation of general-purpose programs.
\newblock In {\em Proceedings of the 25th International Conference on Compiler
  Construction}, pages 121--131. ACM, 2016.

\bibitem{sloss2004arm}
A.~Sloss, D.~Symes, and C.~Wright.
\newblock {\em ARM system developer's guide: designing and optimizing system
  software}.
\newblock Morgan Kaufmann, 2004.

\bibitem{spiliopoulos2011green}
V.~Spiliopoulos, S.~Kaxiras, and G.~Keramidas.
\newblock Green governors: A framework for continuously adaptive dvfs.
\newblock In {\em Green Computing Conference and Workshops (IGCC), 2011
  International}, pages 1--8. IEEE, 2011.

\bibitem{arm2013ambaace}
{Stevens, Ashley}.
\newblock {Introduction to AMBA{\textregistered} 4 ACE{\texttrademark} and
  big.LITTLE{\texttrademark} Processing Technology}.
\newblock Technical report, 2013.

\bibitem{cigar}
{University of Nevada, Reno}.
\newblock {Evolutionary Computing Systems Lab}, 2016.
\newblock [Online] accessed 2016-05-28. Available at
  \url{http://ecsl.cse.unr.edu/}.

\bibitem{van2012scheduling}
K.~Van~Craeynest, A.~Jaleel, L.~Eeckhout, P.~Narvaez, and J.~Emer.
\newblock Scheduling heterogeneous multi-cores through performance impact
  estimation (pie).
\newblock In {\em ACM SIGARCH Computer Architecture News}, volume~40, pages
  213--224. IEEE Computer Society, 2012.

\end{thebibliography}
\end{document}